\documentclass[%
 reprint,
 amsmath,amssymb,
 longbibliography,
 prx
]{revtex4-2}

\usepackage[ruled]{algorithm2e}
\usepackage[noend]{algpseudocode}
\usepackage{type1cm} % used to replace {wasysym}
\usepackage{graphicx}% Include figure files
\usepackage{dcolumn}% Align table columns on decimal point
\usepackage{bm}% bold math
\DeclareMathAlphabet{\mathpzc}{OT1}{pzc}{m}{it}
\begin{document}

%\preprint{APS/123-QED}

\title{Solving multiorbital dynamical mean-field theory using natural orbitals renormalization group}% Force line breaks with \\

\author{Jia-Ming Wang}
\affiliation{Department of Physics, Renmin University of China, Beijing 100872, China}

\author{Yin Chen}
\affiliation{Department of Physics, Renmin University of China, Beijing 100872, China}

\author{Yi-Heng Tian}
\affiliation{Department of Physics, Renmin University of China, Beijing 100872, China}

\author{Rong-Qiang He}\email{rqhe@ruc.edu.cn}
\affiliation{Department of Physics, Renmin University of China, Beijing 100872, China}

\author{Zhong-Yi Lu}\email{zlu@ruc.edu.cn}
\affiliation{Department of Physics, Renmin University of China, Beijing 100872, China}

\date{\today}% It is always \today, but any date may be explicitly specified

\begin{abstract}
  The natural orbitals renormalization group (NORG) has previously been proposed as an efficient numerical method for solving zero-temperature properties of multisite and multiorbital quantum impurity systems. Here, we implement the NORG as an impurity solver for dynamical mean-field theory (DMFT). In comparison with the exact diagonalization method, the NORG method can treat much more bath sites in an impurity model to which the DMFT maps a lattice model and can find accurate zero-temperature Matsubara and low-frequency retarded Green's functions. We demonstrate the effectiveness of this method on a two-orbital Hubbard model on the Bethe lattice and find successfully the orbital selective Mott transition with a Kondo resonance peak in the wide band and two holon-doublon bound state excitation peaks in the narrow band.
\end{abstract}

% provides a new class of research tools \cite{Jordan2015science,Carleo2019RMP}
% contrastive learning of visual representations. \cite{2020SimCLR,wang2022molecular}

%\keywords{Suggested keywords}%Use showkeys class option if keyword display desired
\maketitle

%\tableofcontents

%%%%%%%%%%%%%%%%%%%%%%%%%%%%%%%%%%%%%%%%%%%%%%%%%%%%%%%%%%%%%%%%%%%%%%%%%%%%%%%%%%
%%--------------------------------- Section 1 ----------------------------------%%
%%%%%%%%%%%%%%%%%%%%%%%%%%%%%%%%%%%%%%%%%%%%%%%%%%%%%%%%%%%%%%%%%%%%%%%%%%%%%%%%%%

\section{Introduction}

Strongly correlated materials show a mass of novel phenomena such as high $T_{c}$ superconductivity \cite{Lee2006rmp,Keimer2015nat}, non-Fermi liquid \cite{Stewart2001rmp}, and Mott metal-insulator transition \cite{ Mott1968rmp,Imada1998rmp}. These phenomena have not been understood well to date. The dynamical mean-field theory (DMFT) \cite{Georges1996rmp} helps us understand the electron behaviors in these materials in a non-perturbative way. This theory neglects spatial correlations (assuming the self-energy of a system is local) and is exact in three limits: (1) non-interacting, (2) atomic (infinite interactions), or (3) infinite dimensions. Combined with the density functional theory \cite{Hohenberg1964pr,Jones1989rmp}, the DMFT provides an effective approach to help material first-principles calculations go beyond weak correlations \cite{Kotliar2006rmp,Held2007aip}. However, accurate DMFT results are difficult to obtain in practice \cite{Georges1996rmp}.

The DMFT needs a way to obtain self-energy from a quantum impurity model to solve its self-consistent equations \cite{Georges1996rmp}. Because a quantum impurity model is also a correlated electronic system and hence is also difficult to solve, the DMFT is in great need of efficient impurity solvers. Since the advent of DMFT, people have developed many impurity solvers, such as exact diagonalization \cite{Caffarel1994prl,Granath2012prb,Lu2014prb,Schuler2015prb,Motahari2016prb}, numerical renormalization group (NRG) \cite{Wilson1975RMP,Bulla2008rmp}, density-matrix renormalization group (DMRG) \cite{White1992prl,Schollwock2005rmp,Garc2004prl,Wolf2015prx,Hallberg2015el,Zhu2019prb,Bauernfeind2017prx,Nunez2018fip}, Hirsch-Fye quantum Monte Carlo \cite{Hirsch1986prl}, continuous-time quantum Monte Carlo \cite{Rubtsov2005prb,Werner2006prl,Gull2011rmp}, quantum computing \cite{Bauer2016prx,Sakurai2022prr}, machine learning \cite{Sheridan2021prb}, and some other methods \cite{Pruschke1993prb,Lechermann2007prb,Li2015epjb,Barman2016prb,Go2017prb,Mejuto2019prb,Eidelstein2020prl,Cao2021prb,Li2022prb}. However, none of them works for all cases. The quantum Monte Carlo-type methods are numerically exact but have a sign problem for general systems at low temperatures. The exact diagonalization methods can obtain directly real-frequency spectra but the bath for the impurities can only be discretized into a very small number of bath sites. The NRG focuses on the low-energy range of a system and can not capture the high-energy excitation with sufficient resolution.

In a quantum impurity model, only the impurities have two-body interactions; the other part is a non-interacting electronic bath hybridizing with the impurities. Although a quantum impurity system is also electronically correlated, its correlation is very different from a regular strongly correlated system in that the impurities can only entangle with a finite number of degrees of freedom in the bath, which we call sparse correlation \cite{He2014prb}. Consequently, the ground state is, in some sense, simple \cite{Debertolis2021prb,zheng2018cpl,zheng2020scp} and can be approximately but accurately represented in a very small subspace of the complete Hilbert space. The NRG \cite{Wilson1975RMP,Bulla2008rmp} finds this subspace by selecting many-body basis according to energy (the eigenvalues of the Hamiltonian), the DMRG \cite{White1992prl,Schollwock2005rmp,Garc2004prl,Wolf2015prx,Hallberg2015el,Zhu2019prb,Bauernfeind2017prx,Nunez2018fip} does this by selecting many-body basis according to entanglement (the eigenvalues of the reduced density matrix), while the recently proposed natural orbitals renormalization group (NORG) \cite{He2014prb} does this by selecting many-body basis according to natural orbital occupancies (the eigenvalues of the single-particle density matrix). Here, we focus on the NORG method.

The NORG has been demonstrated as a powerful method for solving quantum impurity models in that it can find the ground state explicitly and the Green's functions accurately for a four-impurity Anderson model \cite{He2014prb} and has resolved the long-standing discrepancy between the NRG and quantum Monte Carlo studies on a two-impurity Kondo problem with up to 1022 bath sites \cite{He2015prb}. Nevertheless, it has never been used as an impurity solver for the DMFT.

In this paper, for the first time, we try to use the NORG as an impurity solver for the DMFT. A half-filled two-orbital Hubbard model on the Bethe lattice is studied as a testbed by the DMFT. The zero-temperature real-frequency spectra are directly obtained from the DMFT mapped quantum impurity model, avoiding the ill-posed analytic continuation. A more realistic density of states (DOS) can be obtained by averaging the DOS's calculated from several quantum impurity models with a different number of bath sites. We find an orbital selective Mott transition (OSMT) when the intraorbital interaction is stronger than the interorbital one, which features a Kondo resonance peak in the wide band and two holon-doublon bound state excitation peaks \cite{Nunez2018prb} in the narrow band. The results are well consistent with those from other studies in the literature \cite{Koga2005prb,Liebsch2005prl,Ferrero2005prb,Medici2009prl,Medici2011prb,Nunez2018prb,Niu2019prb}. The ground state of the quantum impurity model has a low correlation entropy, about ${\rm ln} 5$ in the OSMT regime, which shows that the electronic correlation is sparse and accounts for the high efficiency of the NORG method as an impurity solver for the DMFT.

%%%%%%%%%%%%%%%%%%%%%%%%%%%%%%%%%%%%%%%%%%%%%%%%%%%%%%%%%%%%%%%%%%%%%%%%%%%%%%%%%%
%%--------------------------------- Section 2 ----------------------------------%%
%%%%%%%%%%%%%%%%%%%%%%%%%%%%%%%%%%%%%%%%%%%%%%%%%%%%%%%%%%%%%%%%%%%%%%%%%%%%%%%%%%

\section{Mode and method}

In this section, we introduce briefly the quantum lattice model we will study and the DMFT and the NORG method used to study this model.

\subsection{The quantum lattice model}

We choose the half-filled two-orbital Hubbard model on the Bethe lattice (with an infinite coordination number) as an example of strongly correlated systems to demonstrate the effectiveness of the NORG method. The Hamiltonian of this model is
\begin{equation}
H=-\sum_{\langle ij\rangle l\sigma}t_{l}c_{il\sigma}^{\dagger}c_{jl\sigma}+U\sum_{il}n_{il\uparrow}n_{il\downarrow}+U^{\prime}\sum_{i\sigma\sigma'}n_{i1\sigma}n_{i2\sigma'},\label{eq:lattmodel}
\end{equation}
where $c_{il\sigma}^{\dagger}$ and $c_{il\sigma}$ are the electron creation and annihilation operators for the orbital $l$ on site $i$ with spin $\sigma$ and $n_{il\sigma}=c_{il\sigma}^{\dagger}c_{il\sigma}$. $\langle ij\rangle$ means only the nearest-neighbor hoppings are considered. $t_{l}$ is the hopping integral and $t_{1}>t_{2}$. $U$ and $U'$ are the intra- and inter-orbital Coulomb repulsion strengths, respectively. Because of the infinity coordination number, the DMFT becomes an exact theory for this model and the non-interacting local Green's function has a simple form $G_{l}(z)=\left(z-\sqrt{z^{2}-4t_{l}^{2}}\right)/2t_{l}^{2}$. The orbital-dependent non-interacting half bandwidth $D_{l}=2t_{l}$. We set $D_{1}=1$ as the energy unit. The widths of the two bands are different. Band 1 is the wide band (WB), while band 2 the narrow band (NB). When the interactions are turned on, the model features an OSMT when $\Delta=U-U^{\prime}\ne0$.

\subsection{The DMFT}

The DMFT neglects correlations between different lattice sites (namely only onsite self-energies exist, $\Sigma_{ll^{\prime}}=\delta_{ll^{\prime}}\Sigma_{ll}$) and maps the lattice model to a self-consistently determined quantum impurity model,

\begin{align}
H_{\rm qim}=H_{\text{imp}}+H_{\text{bath}}+H_{\text{hyb}},\label{eq:impmodel}
\end{align}
where $H_{\text{imp}}=U\sum_{l}n_{l\uparrow}n_{l\downarrow}+U^{\prime}\sum_{\sigma\sigma'}n_{1\sigma}n_{2\sigma'}$ is for the impurity site and coincides with the Hamiltonian of a single lattice site of the original lattice model, and assumes $G_{\rm loc}(z) = G_{\rm imp}(z)$, where $G_{\rm loc}(z)$ is the lattice local Green's function while $G_{\rm imp}(z)$ is the impurity Green's function of $H_{\rm qim}$. $H_{\text{bath}}=\sum_{lk\sigma}\epsilon_{lk}b_{lk\sigma}^{\dagger}b_{lk\sigma}+h.c.$ describes an electronic bath for the impurity and $H_{\text{hyb}}=\sum_{lk\sigma}V_{lk}c_{l\sigma}^{\dagger}b_{lk\sigma}+h.c.$ the hybridization between the impurity and the bath. $H_{\text{bath}}$ and $H_{\text{hyb}}$ are subject to equation $G_{\text{0,imp}}^{-1}=\mathcal{G}^{-1}$, where $G_{\text{0,imp}}$ is the impurity non-interaction Green's function and $\mathcal{G}$ the Weiss field for a lattice site.

The bath is discretized and the parameters in $H_{\rm bath}$ and $H_{\rm hyb}$ are determined by fitting $z-\mathcal{G}^{-1}$ with the impurity hybridization function
\begin{equation}
  \Gamma_{\rm imp}(z)=\sum_{k}\frac{|V_{lk}|^{2}}{z-\epsilon_{lk}}\label{eq:hyb_imp}
\end{equation}
(like in the exact diagonalization impurity solver).

The DMFT self-consistent equation for (\ref{eq:lattmodel}) is $\mathcal{G}^{-1}(z)=z-t^{2}G_{\rm imp}(z)$ and is solved usually by an iteration procedure with an initial guessed $\mathcal{G}^{-1}$, where the impurity Green's function $G_{\rm imp}$ of $H_{\rm qim}$ is found by an impurity solver. The DMFT equation can be solved with real or imaginary frequencies. Here, as a preliminary attempt, we do it with imaginary frequencies as most DMFT studies do. After convergence of DMFT iterations, we directly obtain the retarded Green's function from the quantum impurity model.

\subsection{The NORG as an impurity solver}

\begin{figure}[htbp!]
  \includegraphics[width=8.6cm]{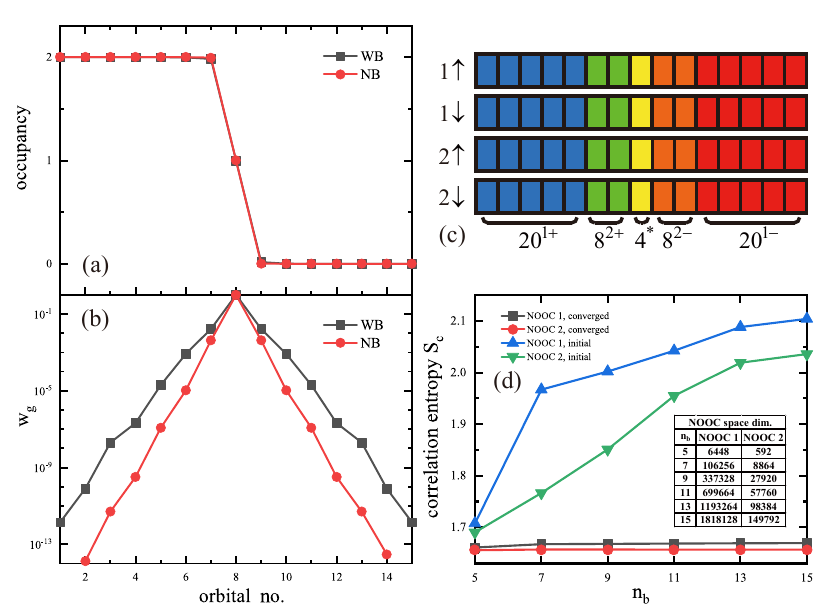}
  \caption{(a) The occupancy number of the natural orbitals found by the NORG for the quantum impurity model (\ref{eq:impmodel}) self-consistently determined in the DMFT for model (\ref{eq:lattmodel}) with $U = 3$, $\Delta = 0.3$, $t_2 = 0.5 t_1$, and $n_{\rm b} = 15$. To save computational cost, the impurity orbitals are not involved in the natural orbital optimization. Most bath natural orbitals are nearly fully occupied or empty, shown more clearly in (b) with $w_g = {\rm min}(n_g, 2 - n_g)$ ($g$ enumerates the bath natural orbitals). The NORG selects many-body basis vectors in the natural orbital basis by a rule that they have to satisfy a natural orbital occupancy constraint (NOOC) to exclude a large number of basis vectors with neglectable contributions in the ground state. A typical NOOC is shown in (c). The natural orbitals are sorted in a descending order according to their occupancy and then divided into five groups. The first and second (fifth and fourth) groups are nearly fully occupied (empty) and allow at most 1 and 2 holes (electrons), respectively, while the middle group is unrestricted and is mostly entangled with the impurity. (d) Correlation entropy $S_{\rm c}$ versus $n_{\rm b}$. NOOC 1 is $(2n_{\rm b}-10)^{1+}8^{2+}4^{*}8^{2-}(2n_{\rm b}-10)^{1-}$ and NOOC 2 is $(2n_{\rm b}-10)^{1+}8^{1+}4^{*}8^{1-}(2n_{\rm b}-10)^{1-}$.}
  \label{fig:nooc-combined}
\end{figure}

For a quantum impurity system, only the impurity has interactions, namely the interaction is {\em sparse}. As a consequence, the correlation in the system is strong and, however, sparse. The ground state ($|\psi\rangle$), when expanded in the natural orbital basis, consists of a very small number of Slater determinants ($x$'s) compared to the size of the whole Hilbert space \cite{He2014prb,Debertolis2021prb}. The natural orbitals are the eigen basis of the single-particle density matrix $D$ and the occupancy numbers of orbitals are extremal in this basis, where $D_{\alpha \beta} = \langle \psi | c_\alpha^\dagger c_\beta |\psi \rangle$. Fig.~\ref{fig:nooc-combined}(a) and (b) is an example.

% Diagonalizing the single-particle density matrix of a many-body wave function, one can obtain the natural orbitals and their electron occupancy number.

The NORG uses this property of a quantum impurity system to choose a subspace with a reduced basis to optimize the wave function. Under the single-particle basis represented by the natural orbitals, some of the Hartree-Fock basis vectors (occupancy configurations or Slater determinants) are discarded. The retained basis vectors satisfy a rule, which we refer to as natural orbital occupancy constraint (NOOC) as described in Fig.~\ref{fig:nooc-combined}(c). The NORG does an exact diagonalization in the reduced Hilbert space (the NOOC space) and finds the natural orbitals and ground state iteratively since they are both unknown at the beginning \cite{He2014prb}.

The sparse correlation in quantum impurity models makes the NORG method very efficient. We can define a quantity, which we refer to as correlation entropy, to measure the sparsity of correlation in a many-body wave function $\psi(x)$,
\begin{equation}
  S_{\rm c}=-\sum_{x}\psi^{2}\left(x\right){\rm log}\psi^{2}\left(x\right).
  \label{eq:sc}
\end{equation}
As shown in Fig.~\ref{fig:nooc-combined}(d), initially the natural orbitals are unknown and the calculated $S_{\rm c}$ is large. When the NORG iteration converges, the natural orbitals are found and $S_{\rm c}$ is actually small and grows very slowly as $n_{\rm b}$ increases and is dominated by the number of degrees of freedom of the impurity so that the NORG has a low computational complexity $O(n_{\rm b}^3)$ with respect to the number of bath sites.

When the ground state is found, we can accurately calculate physical quantities of $H_{\rm qim}$, including Green's functions \cite{He2014prb,He2015prb}. We can calculate real-frequency spectra directly, avoiding the ill-posed analytic continuation problem.

%%%%%%%%%%%%%%%%%%%%%%%%%%%%%%%%%%%%%%%%%%%%%%%%%%%%%%%%%%%%%%%%%%%%%%%%%%%%%%%%%%
%%--------------------------------- Section 3 ----------------------------------%%
%%%%%%%%%%%%%%%%%%%%%%%%%%%%%%%%%%%%%%%%%%%%%%%%%%%%%%%%%%%%%%%%%%%%%%%%%%%%%%%%%%

\section{Results}

To demonstrate the performance of the NORG as a DMFT impurity solver, we employ the DMFT to study the zero-temperature properties of the half-filled two-orbital Hubbard model (\ref{eq:lattmodel}). The NOOC used in the implementation of the NORG is
$$(2n_{\rm b}-10)^{1+}8^{2+}4^{*}8^{2-}(2n_{\rm b}-10)^{1-} ~~~~ (\rm NOOC~1)$$ 
shown in Fig.~\ref{fig:nooc-combined}(c). The number of the retained many-body basis states (the NOOC space dimension) is shown in the inset of Fig.~\ref{fig:nooc-combined}(d).

\subsection{Bath fitting}

\begin{figure}[htbp!]
  \includegraphics[width=8.6cm]{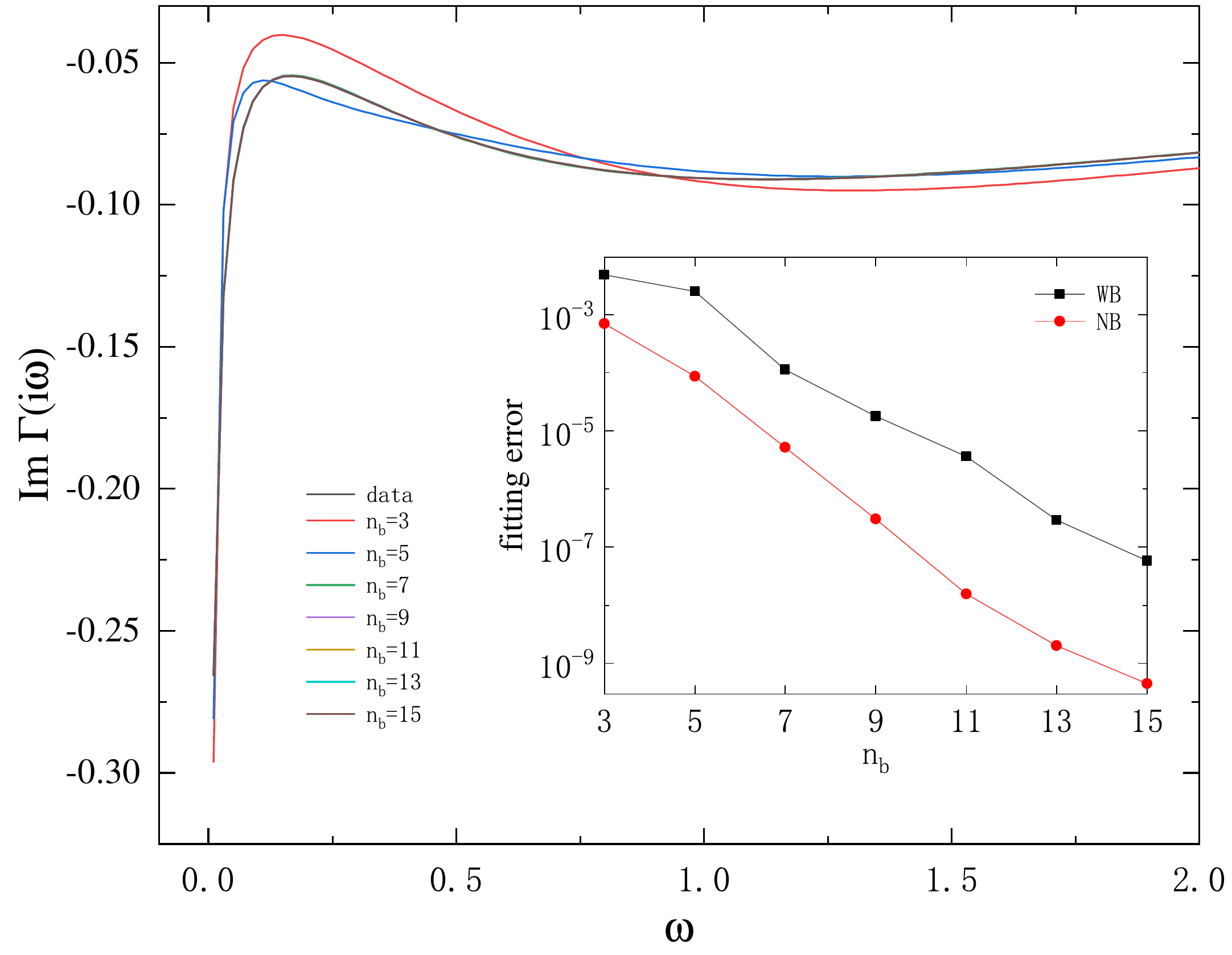}
  \caption{An example of bath fitting. The parameters in $H_{\rm bath}$ and $H_{\rm hyb}$ in the DMFT mapped quantum impurity model (\ref{eq:impmodel}) are determined by fitting $i \omega - \mathcal{G}^{-1}$ with the impurity hybridization function $\Gamma_{\rm imp}(i \omega)$. $U = 3$, $\Delta = 0.3$, $t_2 = 0.5 t_1$. The fitting for the wide band is shown. Inset: the fitting error decreases exponentially as $n_{\rm b}$ increases.}
  \label{fig:fit-hyb-err-u3d0.3wb}
\end{figure}

The parameters in $H_{\rm bath}$ and $H_{\rm hyb}$ in the DMFT mapped quantum impurity model (\ref{eq:impmodel}) are determined by fitting $z-\mathcal{G}^{-1}$ with the impurity hybridization function $\Gamma_{\rm imp}(z)$. We show an example of the fitting in Fig.~\ref{fig:fit-hyb-err-u3d0.3wb}. $n_{\rm b}$ denotes the number of discretized bath sites per impurity orbital. For $n_{\rm b} = 3$ or $5$, the fitting is not well. When $n_{\rm b} \ge 9$, the fitting error is neglectable. In fact, the fitting error decreases exponentially as $n_{\rm b}$ increases, as shown in the inset of Fig.~\ref{fig:fit-hyb-err-u3d0.3wb}. $n_{\rm b} = 15$ makes the fitting error less than $10^{-7}$. So it is unnecessary to use a larger $n_{\rm b}$.

\subsection{Orbital selective Mott transition}

\begin{figure}[htbp!]
  \includegraphics[width=8.6cm]{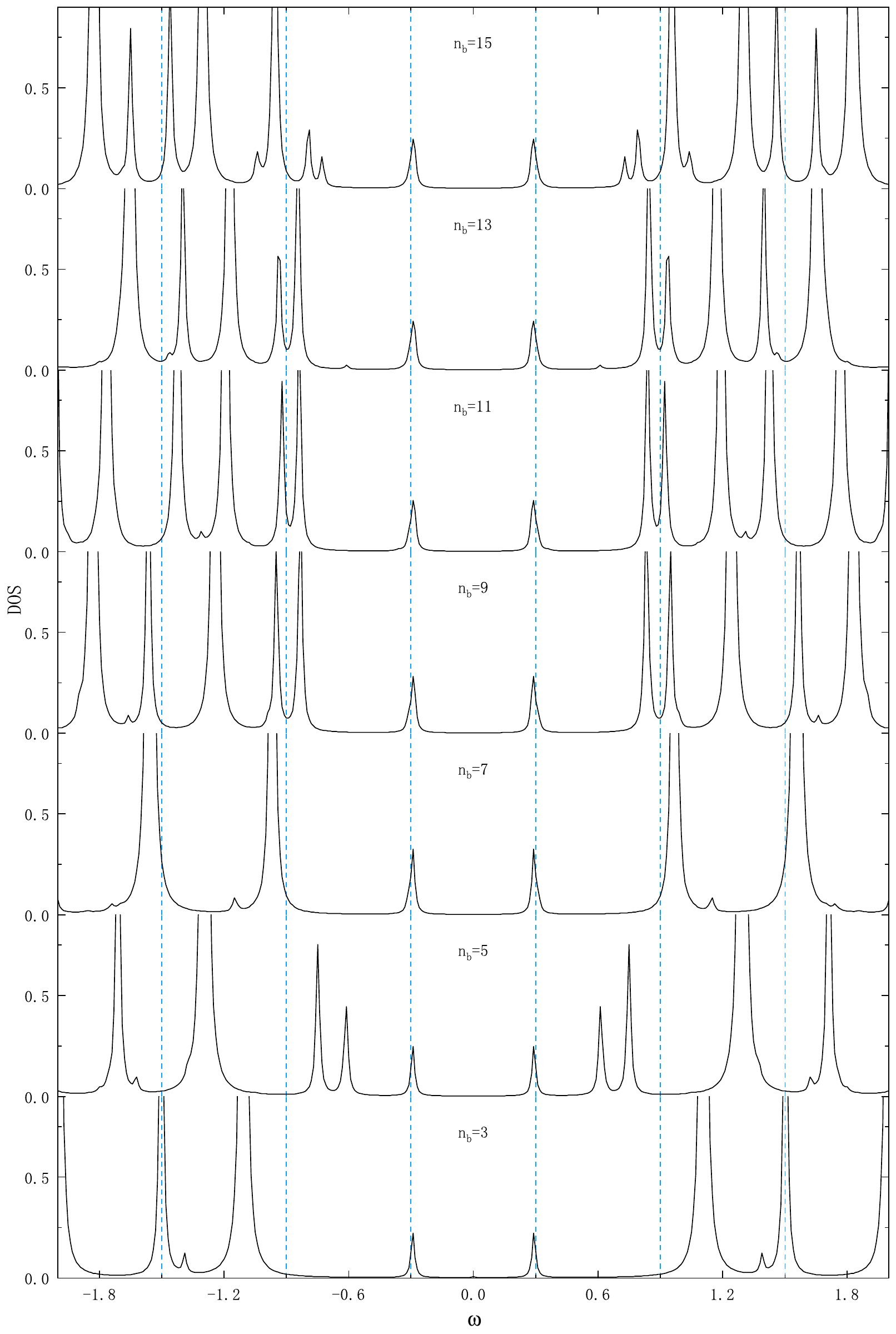}
  \caption{DOS of the narrow band with different $n_{\rm b}$. $U = 3$, $\Delta = 0.3$, $t_2 = 0.5 t_1$. There are holon-doublon bound state excitation peaks at about $\omega = \pm \Delta$.}
  \label{fig:dos-u3d0.3a0b5-15nb}
\end{figure}

The real-frequency spectra are obtained directly from the quantum impurity model by NORG. The density of states (DOS) of the narrow band with different $n_{\rm b}$ are shown in Fig.~\ref{fig:dos-u3d0.3a0b5-15nb}. Because of the finite number of bath sites in the quantum impurity model, the DOS consists with delta peaks. For $n_{\rm b} = 3, 5$, and $7$, the positions of the peaks fluctuate arbitrarily, which should be attributed to the poor bath fitting (Fig.~\ref{fig:fit-hyb-err-u3d0.3wb}). In contrast, for larger $n_{\rm b}$, the positions of the peaks stay relatively stable and the upper and lower Hubbard bands are clearly seen. Then, it is reasonable to average the DOS's of $n_{\rm b} = 9-15$ to obtain a more realistic DOS without losing true information.

\begin{figure}[htbp!]
  \includegraphics[width=8.6cm]{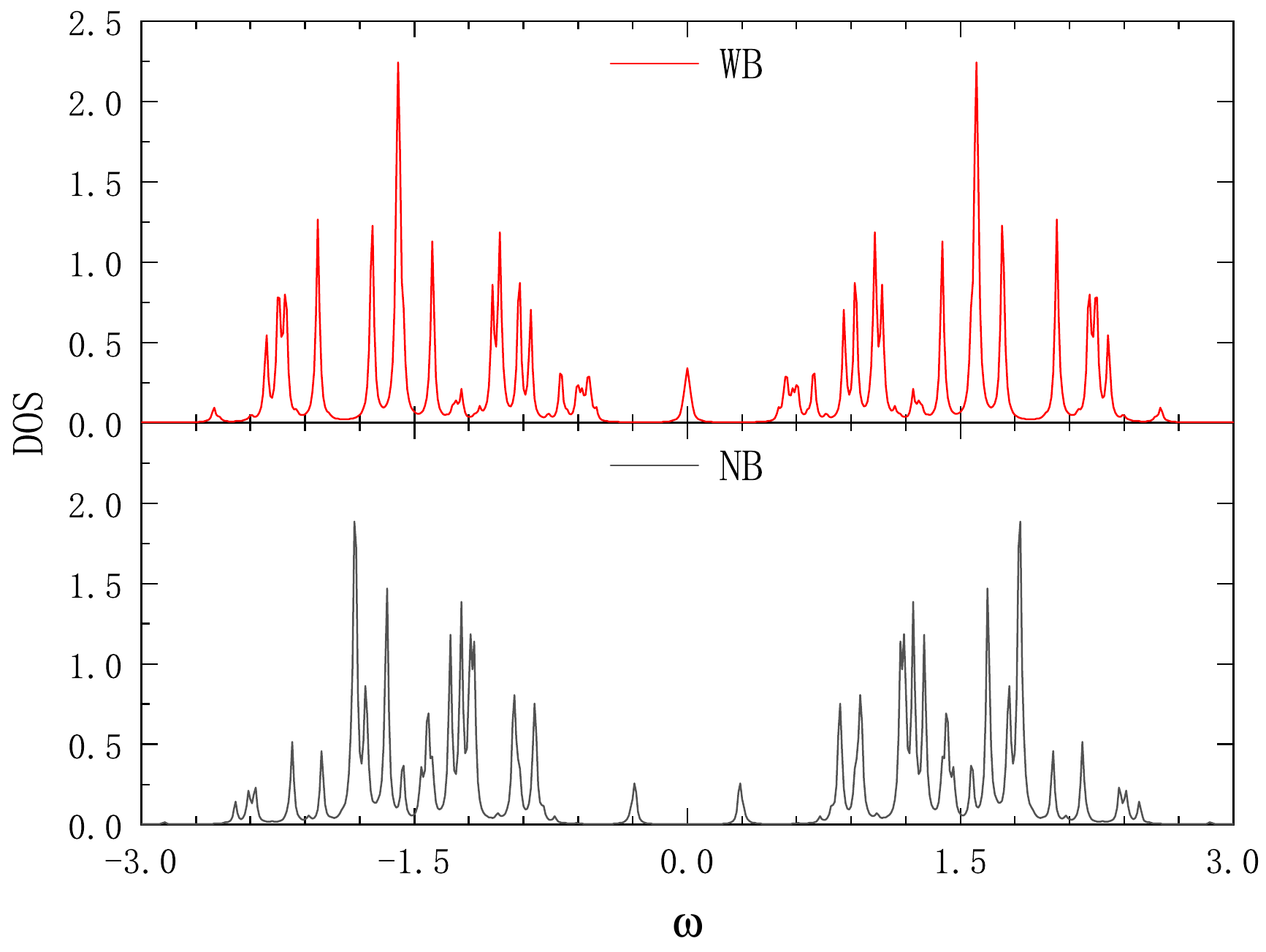}
  \caption{Averaged DOS of $n_{\rm b} = 9, 11, 13, 15$. $U = 3$, $\Delta = 0.3$, $t_2 = 0.5 t_1$. There are holon-doublon bound state excitation peaks at about $\omega = \pm \Delta$ in the narrow band and a Kondo resonance peak at $\omega = 0$.}
  \label{fig:dos-u3d0.3a0b9-15avg}
\end{figure}

We show the averaged DOS in Fig.~\ref{fig:dos-u3d0.3a0b9-15avg} for $U = 3$, $\Delta = 0.3$, $t_2 = 0.5 t_1$. The wide band is metallic while the narrow band is insulating, which indicates an OSMT. There is a Kondo resonance peak in the wide band at zero energy, while there are robust excitation peaks in the narrow band at about $\omega = \pm \Delta$. As $U$ increases, the two kinds of peaks will vanish simultaneously, which implies that they are related to each other or share the same origin. Ref.~\cite{Nunez2018prb} shows that the quasiparticle excitations in the narrow band at $\omega = \pm \Delta$ are holon-doublon bound states. The peak with particle (hole) excitation is related to a holon-doublon bound state with a holon (doublon) state in the wide band and a doublon (holon) state in the narrow band. Because of the existence of the inter-orbital interaction $U^\prime$, the energy of the holon-doublon bound states is only about $\Delta$ higher than the ground state energy. Because the wide band is metallic, there will be holons and doublons in the wide band while there are only singly occupied states in the narrow band in the ground state. The particle (hole) excitation in the narrow band will have overlap with the holon-doublon bound states. The Kondo peak in the wide band also relies on the fact that the wide band is metallic and holons and doublons are allowed to exist. So the holon-doublon bound state excitation peaks in the narrow band share the same origin with the central coherent peak in the wide band. Actually, a careful inspection shows that the spectral weight of the Kondo peak is the same as the total spectral weight of the two quasiparticle peaks in the narrow band.

\subsection{Absence of orbital selective Mott transition}

\begin{figure}[htbp!]
  \includegraphics[width=8.6cm]{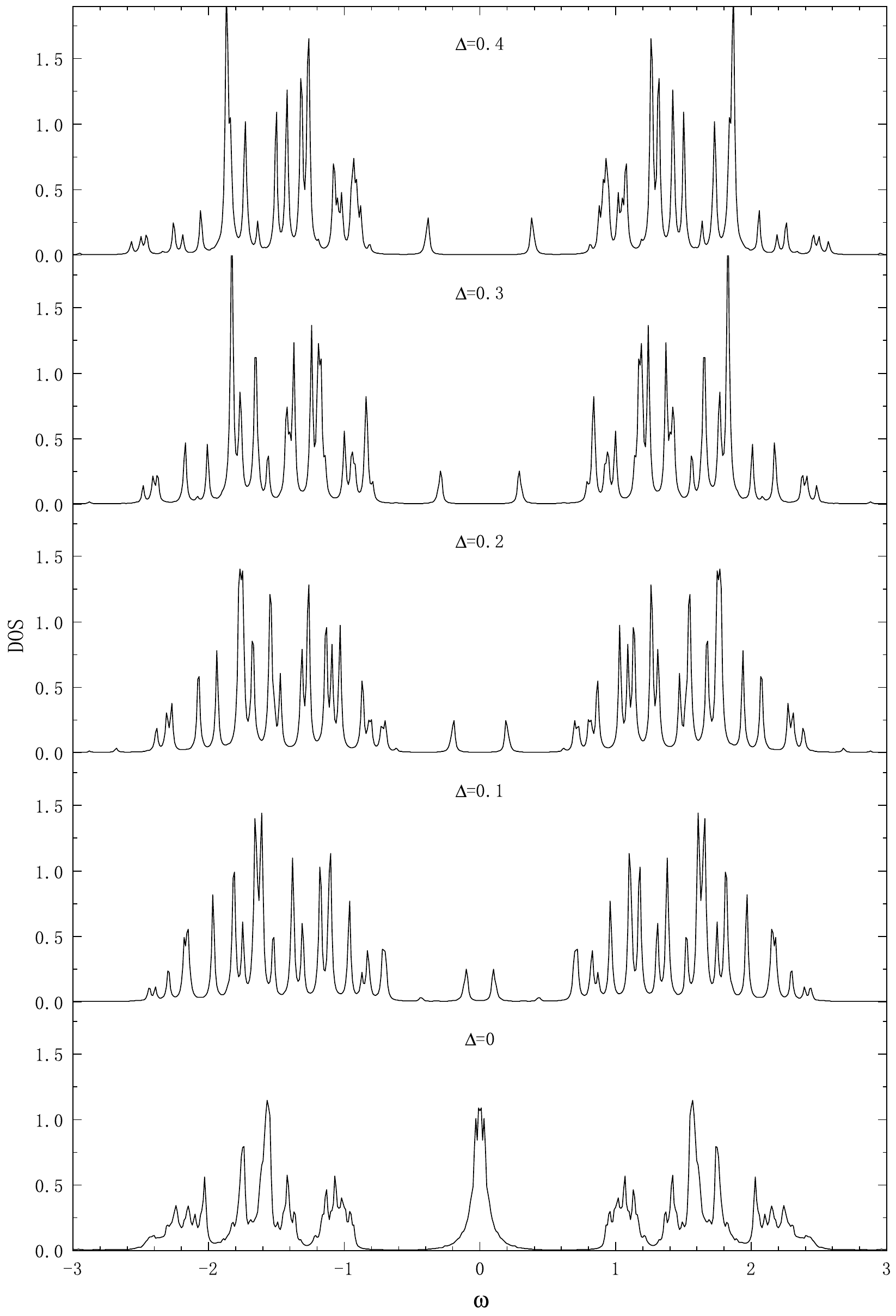}
  \caption{Averaged DOS of the narrow band with different $\Delta$. $U = 3$, $t_2 = 0.5 t_1$. As $\Delta$ decreases, the two holon-doublon bound state excitation peaks in the narrow band moves gradually to zero frequency while the central peak in the wide band remains (now shown).}
  \label{fig:dos-u3d0-0.4}
\end{figure}

\begin{figure}[htbp!]
  \includegraphics[width=8.6cm]{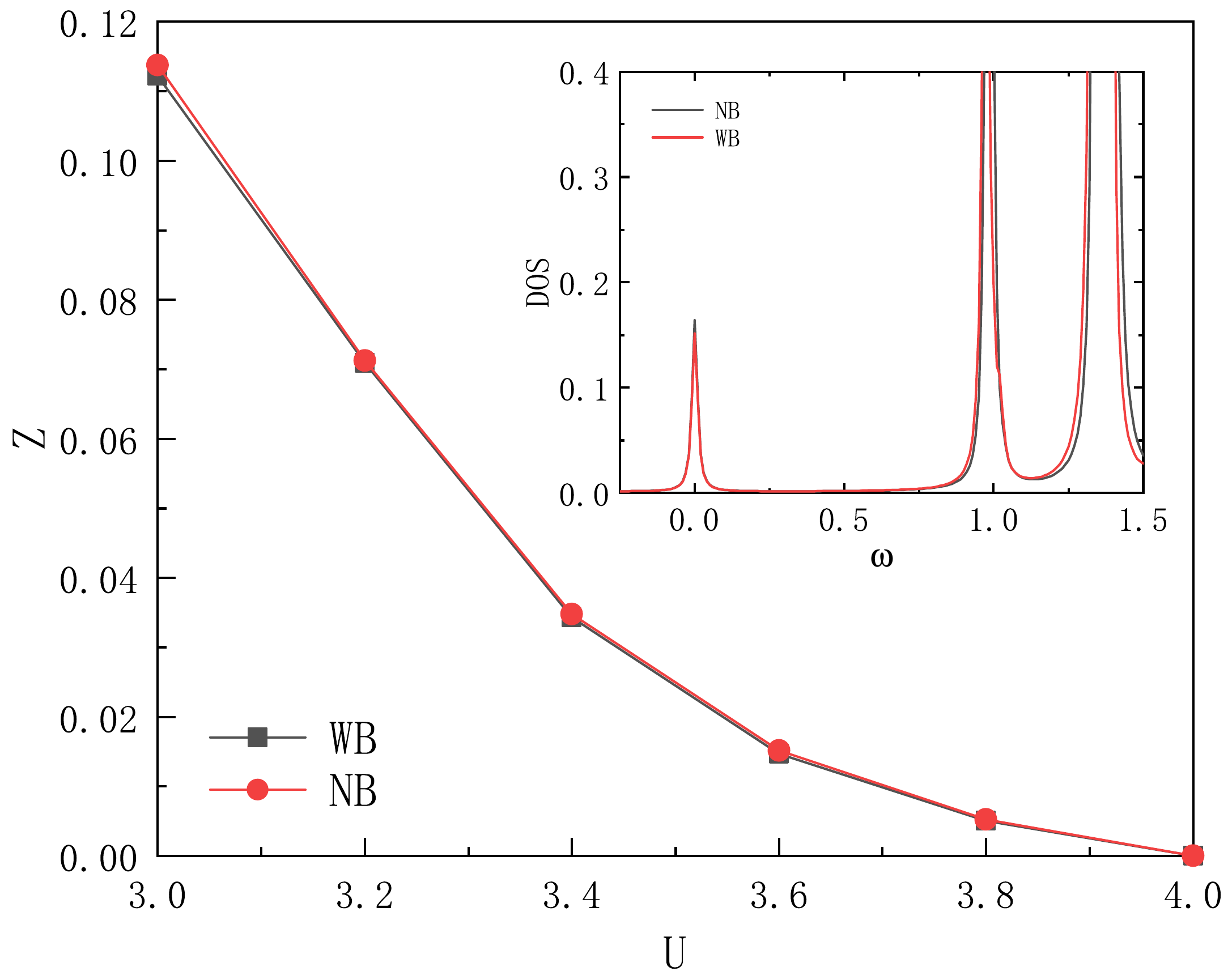}
  \caption{Quasiparticle weight versus the intra-orbital interaction $U$. $\Delta = 0$, $t_2 = 0.5 t_1$, $n_{\rm b} = 11$. The inset shows the corresponding DOS's with $U = 3.8$. The quasiparticle weights for the central peaks in the two different bands are always equal and the OSMT is absent.}
  \label{fig:z-d0b11-inset-u3.8}
\end{figure}

As shown in Fig.~\ref{fig:dos-u3d0-0.4}, as $\Delta$ decreases, the two holon-doublon bound state excitation peaks in the narrow band move gradually to zero frequency. When $\Delta = 0$, the two peaks merge at zero frequency and the narrow band becomes metallic. For this symmetric interaction ($U = U^\prime$), although both the bands are metallic and have central peaks, the natures of the peaks are different. The central peak in the wide band is a Kondo resonance, while that in the narrow band is a holon-doublon bound state excitation. They are `locked' to each other and disappear simultaneously when increasing $U$. As shown in Fig.~\ref{fig:z-d0b11-inset-u3.8}, the quasiparticle weights ($Z=[1-\frac{\partial Re\Sigma(\omega)}{\partial \omega}|_{\omega=0}]^{-1}$) for the two central peaks are always equal. The OSMT is absent and we only see a simultaneous Mott transition.

%%%%%%%%%%%%%%%%%%%%%%%%%%%%%%%%%%%%%%%%%%%%%%%%%%%%%%%%%%%%%%%%%%%%%%%%%%%%%%%%%%
%%--------------------------------- Section 4 ----------------------------------%%
%%%%%%%%%%%%%%%%%%%%%%%%%%%%%%%%%%%%%%%%%%%%%%%%%%%%%%%%%%%%%%%%%%%%%%%%%%%%%%%%%%

\section{Conclusion and outlook}

In summary, through studying a half-filled two-orbital Hubbard model, the natural orbitals renormalization group (NORG) has been demonstrated as an efficient impurity solver for dynamical mean-field theory (DMFT). We have calculated the zero-temperature real-frequency spectra directly from the DMFT mapped multiorbital quantum impurity model and found an orbital selective Mott transition and interband holon-doublon bound states when the intraorbital interaction is stronger than the interorbital one. The results are well consistent with other studies in the literature \cite{Koga2005prb,Liebsch2005prl,Ferrero2005prb,Medici2009prl,Medici2011prb,Nunez2018prb,Niu2019prb}.

The NORG complements other impurity solvers like exact diagonalization and quantum Monte Carlo and is worth exploring more. The bath for the DMFT mapped quantum impurity model has been discretized to a set of bath sites like in the exact diagonalization impurity solver. The parameters for the quantum impurity model are obtained by fitting. 9 bath sites per impurity orbital can make the fitting error neglectable when the DMFT self-consistent equations are solved on the imaginary frequencies. Much more bath sites can be treated by the NORG and this will take effect when the DMFT self-consistent equations are solved on the real frequencies and increase the resolution for high-energy excitations. Much more bath sites will constitute a big bath, which will allow the electron density to be well controlled and hence a more accurate DMFT study of systems without particle-hole symmetry and of superconductivity with doping.

\begin{acknowledgments}
  This work was supported by National Natural Science Foundation of China (Grants No. 11874421 and No. 11934020). Computational resources were provided by Physical Laboratory of High Performance Computing in RUC.
\end{acknowledgments}

%\nocite{*}
\bibliography{imhf2b}% Produces the bibliography via BibTeX.

\end{document}